\documentclass{article} 
\usepackage{epsfig} 
\usepackage{cite} 
\def\H{\hbox{H}} 
\def\Hp{\hbox{H}_+} 
\def\Hm{\hbox{H}_-} 
\def\e{\hbox{e}} 
\def\d{\hbox{d}} 
\def\f{\hbox{f}} 
\def\S{\hbox{S}} 
\def\Li{\hbox{Li}} 
\def\ln{\hbox{ln}} 
\def\hpl{HPL} 
\def\ie{{\it i.e.}} 
\def\iep{i\epsilon} 

\begin{document} 
\unitlength1cm 
\begin{titlepage} 
\vspace*{-1cm} 
\begin{flushright} 
CERN-TH/2001-188\\ 
hep-ph/0107173\\ 
July 2001 
\end{flushright} 
\vskip 3.5cm 
\renewcommand{\topfraction}{0.9}
\renewcommand{\textfraction}{0.0}

\begin{center} 
\boldmath 
{\Large\bf Numerical Evaluation of Harmonic Polylogarithms}\unboldmath 
\vskip 1.cm 
{\large  T.~Gehrmann}$^a$ and {\large E.~Remiddi}$^b$ 
\vskip .7cm 
{\it $^a$ Theory Division, CERN, CH-1211 Geneva 23, Switzerland} 
\vskip .4cm 
{\it $^b$ Dipartimento di Fisica, 
    Universit\`{a} di Bologna and INFN, Sezione di 
    Bologna,  I-40126 Bologna, Italy} 
\end{center} 
\vskip 2.6cm 

\begin{abstract} 
Harmonic polylogarithms $\H(\vec{a};x)$, a generalization of 
Nielsen's polylogarithms ${\rm S}_{n,p}(x)$, appear frequently in analytic 
calculations of radiative corrections in quantum field theory. 
We present an algorithm for the numerical evaluation of harmonic 
polylogarithms of arbitrary real argument. This algorithm is implemented 
into a {\tt FORTRAN} subroutine {\tt hplog} to compute harmonic 
polylogarithms up to weight 4.
\end{abstract} 
\vfill 
\end{titlepage} 
\newpage 

{\bf\large PROGRAM SUMMARY}
\vspace{4mm}

\begin{sloppypar}
\noindent   {\em Title of program\/}: {\tt hplog} \\[2mm]
   {\em Version\/}: 1.0 \\[2mm]
   {\em Release\/}: 1  \\[2mm]
   {\em Catalogue number\/}: \\[2mm]
   {\em Program obtained from\/}:
   {\tt Thomas.Gehrmann@cern.ch, Ettore.Remiddi@bo.infn.it} \\[2mm]
   {\em E-mail\/}: {\tt Thomas.Gehrmann@cern.ch, Ettore.Remiddi@bo.infn.it} \\[2mm]
   {\em Licensing provisions\/}: none \\[2mm]
   {\em Computers\/}: all \\[2mm]
   {\em Operating system\/}: all \\[2mm]
   {\em Program language\/}: {\tt FORTRAN77     } \\[2mm]
   {\em Memory required to execute\/}: Size: 516k \\[2mm]
   {\em No.\ of lines in distributed program\/}: 3271 \\[2mm]
   {\em Other programs called\/}: none \\[2mm]
   {\em External files needed\/}: none \\[2mm]
   {\em Nature of the physical problem\/}: 
        Numerical evaluation of the harmonic polylogarithms up to weight 4 
        for arbitrary real argument.  These functions are emerging in 
        Feynman graph integrals involving more than one mass scale. \\[2mm]
   {\em Method of solution\/}: 
        For small values of the argument: series representation; other 
        values of the argument: transformation formulae. \\[2mm] 
   {\em Restrictions on complexity of the problem\/}: limited up 
        to HPL of weight 4, the algorithms used here can be extended to 
        higher weights without modification. \\[2mm]
   {\em Typical running time\/}:
        On average 0.3 ms for the evaluation of all harmonic polylogarithms up 
        to weight 4 on a Pentium III/600 MHz Linux PC. 
\end{sloppypar}

\newpage

{\bf\large LONG WRITE-UP}
\vspace{4mm}

\renewcommand{\theequation}{\mbox{\arabic{section}.\arabic{equation}}} 

\section{Introduction} 
\label{sec:int} 
\setcounter{equation}{0}

The Euler Dilogarithm ${\rm Li}_2(x)$, and its generalizations, 
Nielsen's polylogarithms~\cite{Nielsen}, have been playing a central 
role in the analytic evaluation of integrals arising in 
perturbative quantum 
field theory. A reliable and 
widely used numerical representation of these functions 
({\tt GPLOG})~\cite{bit}
has been available for already thirty years. Going to higher 
orders in perturbation theory, it was realized recently that 
Nielsen's polylogarithms are insufficient to evaluate all integrals 
appearing in Feynman graphs at two loops and beyond. This limitation 
can only be overcome by the introduction of further generalizations 
of Nielsen's polylogarithms, the Harmonic Polylogarithms (HPLs). 
The HPLs, introduced in \cite{hpl}, together with their ``2-dimensional" 
extensions~\cite{doublebox}, are already now playing a 
central role in the analytic evaluation of Feynman graph 
integrals~\cite{doublebox,moch}. HPLs appear also as inverse Mellin 
transformations of harmonic sums, 
which were investigated and implemented numerically in~\cite{blumlein}.

Let 
us recall here two key features of an analytical calculation. The first is 
to express everything in terms of mathematical quantities whose properties 
are known (so that in particular the final formulae can be written in a unique 
way, which does not hide cancellations between the various terms involved). 
The second is to allow the fast and precise numerical evaluation of all 
the mathematical quantities introduced. Obviously, such fast and precise 
evaluation relies on the knowledge of the analytical properties. 

In this paper, somewhat as a continuation of \cite{bit} dealing with 
the numerical evaluation of Nielsen's polylogarithms~\cite{Nielsen} 
(a subset of the HPLs), we shortly review the analytical properties 
of the HPLs  and then show how they can be used for writing a {\tt FORTRAN} 
code which evaluates the HPLs  up to weight 4 (see Section~\ref{sec:def} 
below for the definition of the weight; 4 is the maximuum weight 
required in the calculations of \cite{doublebox}) with absolute 
precision better than $3\times 10^{-15}$ (\ie \ standard double precision) 
with a 
few dozens of elementary arithmetic operations per function (just a 
single dozen in the most favourable case). 
Given the large number (81) of HPLs  of weight 4 and the many algebraic 
relations among them, our {\tt FORTRAN} routine evaluates the whole set of 
all the HPLs  up to the required weight -- at variance with 
{\tt GPLOG}~\cite{bit}, which evaluates separately the various Nielsen's 
polylogarithms. 
\par 
The plan of the paper is as follows. 
Section~\ref{sec:def} recalls the definitions of the HPLs. Their
algebraic properties are discussed in 
Section~\ref{sec:alg}, where we show how to use these properties 
for separating the functions into reducible and irreducible ones. 
Section~\ref{sec:anal} studies the analytic properties which allow to 
perform converging power series expansions and to accelerate their 
convergence. Relations between HPLs  for different ranges of the arguments 
are derived in Section~\ref{sec:ids}. 
Section~\ref{sec:numer} explains  how the properties recalled above 
are used to implement the HPLs  into a {\tt FORTRAN} subroutine 
{\tt hplog}, and Section~\ref{sec:checks} 
how the correct implementation is checked. 
Finally, we describe the usage of the subroutine {\tt hplog} in 
Section~\ref{sec:code} and provide a few numerical examples in 
Section~\ref{sec:plots}.

\section{Definitions} 
\label{sec:def} 
\setcounter{equation}{0} 
The Harmonic Polylogarithms (HPLs), introduced in \cite{hpl}, are 
one-variable functions $ \H(\vec{a};x) $ depending, besides the argument 
$x$, on a set of indices, grouped for convenience into the vector 
$\vec{a}$, whose components can take one of the three values $(1,0,-1)$ 
and whose number is the weight $w$ of the \hpl. More explicitly, the three 
HPLs  with $w=1$ are defined as 
\begin{eqnarray} 
  \H(1;x) &=& \int_0^x \frac{\d x'}{1-x'} = - \ln(1-x) \ , \nonumber\\ 
  \H(0;x) &=& \ln x \ ,          \nonumber\\ 
  \H(-1;x) &=& \int_0^x \frac{\d x'}{1+x'} = \ln(1+x) \ ; 
\label{eq:defineh1} 
\end{eqnarray} 
their derivatives can be written as 
\begin{equation} 
  \frac{\d }{\d x} \H(a;x) = \f(a;x) \ , \hskip 1cm a=1,0,-1 
\label{eq:derive1} 
\end{equation} 
where the 3 rational fractions $f(a;x)$ are given by 
\begin{eqnarray} 
   \f(1;x) &=& \frac{1}{1-x} \ , \nonumber\\ 
   \f(0;x) &=& \frac{1}{x} \ , \nonumber\\ 
   \f(-1;x) &=& \frac{1}{1+x} \ . 
\label{eq:definef} 
\end{eqnarray} 
For weight $w$ larger than 1, write $ \vec{a} = (a, \vec b) $, where 
$a$ is the leftmost component of  $ \vec{a} $ and $\vec b $ stands 
for the vector of the remaining $(w-1)$ components. The harmonic 
polylogarithms of weight $w$ are then defined as follows: 
if all the $w$ components of $\vec a$ take the value 0, $\vec a$ is said 
to take the value $\vec 0_w$ and 
\begin{equation} 
\H(\vec{0}_w;x) = \frac{1}{w!} \ln^w{x} \ , 
\label{eq:defh0} 
\end{equation} 
while, if $\vec{a} \neq \vec{0}_w$ 
\begin{equation} 
\H(\vec{a};x) = \int_0^x \d x' \ \f(a;x') \ \H(\vec{b};x') \ . 
\label{eq:defn0} 
\end{equation} 
In any case the derivatives can be written in the compact form 
\begin{equation} 
\frac{\d }{\d x} \H(\vec{a};x) = \f(a;x) \H(\vec{b};x) \ , 
\label{eq:derive} 
\end{equation} 
where, again, $a$ is the leftmost component of $ \vec a $ and 
$ \vec b $ stands for the remaining $(w-1)$ components. 
\par 
It is immediate to see, from the very definition Eq.\ (\ref{eq:defn0}), that 
there are $3^w$ HPLs  of weight $w$, and that they are linearly 
independent. The HPLs  are generalizations of Nielsen's 
polylogarithms~\cite{Nielsen}. The function $ \S_{n,p}(x) $, in 
Nielsen's notation, is equal to the {\hpl} whose first $n$ indices are all 
equal to 0 and the remaining $p$ indices all equal to 1:
\begin{equation}
   \S_{n,p}(x) = \H(\vec{0}_{n},\vec{1}_p;x) \; ; 
\end{equation} 
in particular the Euler polylogarithms $ \Li_n(x) = \S_{n-1,1}(x) $ 
correspond to 
\begin{equation}
   \Li_n(x) = \H(\vec{0}_{n-1},1;x) \; .
\end{equation} 

All relations between HPLs  which will be used for their 
numerical evaluation are easily derived by using recursively the 
definition (\ref{eq:defn0}) and the related differentiation formula 
(\ref{eq:derive}). 

\section{The algebra and the reduction equations} 
\label{sec:alg} 
\setcounter{equation}{0} 
As shown in \cite{hpl}, the product of two HPLs  of a same argument $x$ 
and weights $p, q$ can be expressed as a combination of HPLs  of that 
argument and weight $r=p+q$, according to the product identity 
\begin{eqnarray} 
 \H(\vec{p};x)\H(\vec{q};x) & = & 
  \sum_{\vec{r} = \vec{p}\uplus \vec{q}} \H(\vec{r};x) \; , 
\label{eq:halgebra} \end{eqnarray} 
where $\vec p, \vec q$ stand for the $p$ and $q$ components of the indices 
of the two HPLs, while $\vec{p}\uplus \vec{q}$ represents all mergers of 
$\vec{p}$ and $\vec{q}$ into the vector $\vec{r}$ with $r$ components, 
in which the relative orders of the elements of $\vec{p}$ and $\vec{q}$ 
are preserved. 
\par 
The explicit formulae relevant up to weight 4 are 
\begin{equation} 
   \H(a;x) \; \H(b;x) =  \H(a,b;x) + \H(b,a;x) \ , 
\label{eq:alg0} 
\end{equation} 
\begin{eqnarray} 
   \H(a;x) \; \H(b,c;x) &=&  \H(a,b,c;x) + \H(b,a,c;x) + \H(b,c,a;x) 
                             \ , \nonumber\\ 
   \H(a;x) \; \H(b,c,d;x) &=&  \H(a,b,c,d;x) + \H(b,a,c,d;x) + \H(b,c,a,d;x) 
                              + \H(b,c,d,a;x) \ , 
\label{eq:alg1} 
\end{eqnarray} 
and 
\begin{eqnarray} 
  \H(a,b;x) \; \H(c,d;x) &=&  \H(a,b,c,d;x) + \H(a,c,b,d;x) + \H(a,c,d,b;x) 
                              \nonumber\\
                         &+& \H(c,a,b,d;x)  + \H(c,a,d,b;x) + \H(c,d,a,b;x) 
                             \ , 
\label{eq:alg2} 
\end{eqnarray} 
where $a,b,c,d$ are indices taking any of the values $(1,0,-1)$. 
The formulae can be easily verified, one at a time, 
by observing that they are true at some specific point (such as $x=0$, 
where all the HPLs  vanish except in the otherwise trivial case in which 
all the indices are equal to 0), then taking the $x$-derivatives of the 
two sides according to Eq.\ (\ref{eq:derive}) and checking that they are equal 
(using when needed the previously established lower-weight formulae). 

Another class of identities is obtained by integrating (\ref{eq:defh0}) 
by parts. These integration-by-parts (IBP) identities read:
\begin{eqnarray}
H(m_1,\ldots,m_q;x) &=&  H(m_1;x)H(m_2,\ldots,m_q;x)
                        -H(m_2,m_1;x)H(m_3,\ldots,m_q;x) \nonumber \\
&& + \ldots + (-1)^{q+1} H(m_q,\ldots,m_1;x)\;.
\label{eq:ibp}
\end{eqnarray} 
These identities are not fully linearly 
independent from the product identities.

By using Eqs.(\ref{eq:alg0}) for 
all possible independent values of the 
indices $a,b$ one obtains 6 independent relations between the 9 HPLs  
of weight 2 and the products of 2 HPLs  of weight 1; those relations 
can be used for expressing 6 of the HPLs  of weight 2 in terms of 3 HPLs  
of weight 2 and products of 2 HPLs  of weight 1. 
The choice of the 3 HPLs  (referred to, in this context, as 
``irreducible") is by no means unique; by choosing as irreducible HPLs  
of weight 2 the 3 functions $ \H(0,1;x), \H(0,-1;x) $ and $\H(-1,1;x) $, 
the reduction equations expressing the 6 ``reducible" HPLs  of weight 2 
in terms of the irreducible HPLs  read 
\begin{eqnarray} 
   \H(1,1;x) &=& \frac{1}{2} \H(1;x)\;\H(1;x) \ , \nonumber\\ 
   \H(1,0;x) &=& \H(1;x)\;\H(0;x) - \H(0,1;x) \ , \nonumber\\ 
   \H(1,-1;x) &=& \H(1;x)\;\H(-1;x) - \H(-1,1;x) \ , \nonumber\\ 
   \H(0,0;x) &=& \frac{1}{2} \H(0;x)\;\H(0;x) \ , \nonumber\\ 
   \H(-1,0;x) &=& \H(-1;x)\;\H(0;x) - \H(0,-1;x) \ , \nonumber\\ 
   \H(-1,-1;x) &=& \frac{1}{2} \H(-1;x)\;\H(-1;x) \ . 
\label{eq:red2} 
\end{eqnarray} 
Similarly, at weight 3 one has 27 HPLs  and 19 independent product 
and integration-by-parts
identities, expressing 19 reducible HPLs  in terms of 8 irreducible ones; 
at weight 4 there are 81 HPLs, 63 independent identities, 
and correspondingly 63 reducible and 18 irreducible HPLs. 

\section{The analyticity properties} 
\label{sec:anal} 
\setcounter{equation}{0} 
At weight 1, the HPLs  are just logarithms; with the standard conventions 
for the specification of the logarithm, 
it is immediately seen that $\H(1;x)$ has a cut along the real axis 
from $x=1$ to $x \to +\infty$, $\H(0;x)$ from $x=0$ to $x \to -\infty$ and 
$\H(-1;x)$ from $x=-1$ to $x\to -\infty$. With the usual $+\iep$ prescription 
the corresponding complex values along the whole real axis are given by 
\begin{eqnarray} 
   \H(1;x+\iep)  &=& - \ln(|1-x|) + i\pi\theta(x-1) \ , \nonumber\\ 
   \H(0;x+\iep)  &=&  \ln(|x|) + i\pi\theta(-x) \ ,     \nonumber\\ 
   \H(-1;x+\iep) &=&  \ln(|1+x|) + i\pi\theta(-x-1) \ . 
\label{eq:cmp1} 
\end{eqnarray} 
In particular $ \H(1;x)$ and $\H(-1;x) $ are analytic at $x=0$ 
and can therefore be expanded in series of powers of $x$ around $x=0$, 
the first term of the expansion being of order $x$. \par 
At weight 2, let us start from the HPLs  whose rightmost (trailing) 
index is 1, 
$$ \H(a,1;x) = \int_0^x dt\; \f(a,t) \; \H(1;t) \ , \hskip1cm a=1,0,-1 \ .$$ 
For $a=1$, $H(1,1;x)$ has the same cut from $x=1$ to 
$x=+\infty$ as $\H(1;x)$. For $a=0$, by recalling that $\H(1;t)$ can be 
expanded in powers of $t$ at $t=0$, and that the first term of the expansion 
is $t$ ({\ie} the constant term vanishes), one finds that also 
$$ \H(0,1;x) = \int_0^x \frac{dt}{t} \; \H(1;t) $$ 
shares the same analyticity properties as $\H(1;x)$, namely 
it has the same cut from $x=1$ to $x=+\infty$ and it can also be expanded 
in powers of $x$ around $x=0$, the first power of the expansion 
being $x$ ({\ie} the constant term vanishes). For $a=-1$, finally, 
one sees that 
$$ \H(-1,1;x) = \int_0^x \frac{dt}{1+t} \; \H(1;t) $$ 
can again be expanded in $x$ at $x=0$ (the first power of the expansion 
being in this case $x^2$); but besides the right cut $1\leq x<+\infty$ 
implied by the presence of the $ \H(1;t) $ in the definition, 
$ \H(-1,1;x) $ has also the left cut $ -\infty < x \leq -1 $ due to the 
$ 1/(1+t) $ fraction. The two cuts can be easily separated by writing 
\begin{eqnarray} 
   \H(-1,1;x)   &=& \Hp(-1,1;x) + \Hm(-1,1;x) \ , \nonumber\\ 
   \Hp(-1,1;x) &=& \int_0^x \frac{dt}{1+t} \left[ \H(1;t) 
                           - \H(1;-1) \right] \ , \nonumber\\ 
   \Hm(-1,1;x) &=& \int_0^x \frac{dt}{1+t} \left[ \H(1;-1) \right] \ , 
\label{eq:split-11} 
\end{eqnarray} 
where $ \Hp(-1,1;x) $ has only the right cut $1\leq x<+\infty$ and 
$\Hm(-1,1;x) $ only the left cut $ -\infty < x \leq -1 $, (and both admit an 
expansion in $x$ at $x=0$ whose constant term vanishes). Note that 
in the above example the value of $ \H(1;-1)=-\ln 2 $ is finite, as 
$\H(1;x) $ is regular at $x=-1$. 
 
The same discussion applies to the functions $\H(a,-1;x), a=1,0,-1$; 
$\H(-1,-1;x)$ and $\H(0,-1;x)$ share the same analyticity properties 
as $\H(-1;x)$, while 
$$\H(1,-1;x) = \int_0^x \frac{dt}{1-t} \H(-1;t) $$ 
develops, besides the left cut $-\infty<x\leq -1$ implied by $\H(-1;t)$, 
the right cut $1\leq x<+\infty$ due to the 
fraction $ 1/(1-t) $ in the definition, for which one can write:
\begin{eqnarray} 
   \H(1,-1;x)   &=& \Hp(1,-1;x) + \Hm(1,-1;x) \ , \nonumber\\ 
   \Hp(1,-1;x) &=& \int_0^x \frac{dt}{1-t} \H(-1;1) \ , 
                                                  \nonumber\\ 
   \Hm(1,-1;x) &=& \int_0^x \frac{dt}{1-t} \left[ \H(-1;t) 
                                                - \H(-1;1) \right] \ , 
\label{eq:split1-1} 
\end{eqnarray} 
where $\H(-1;1)=\ln 2$ is well defined as $\H(-1;x)$ 
is regular at $x=-1$. 
\par 
The procedure might be extended to cover also the HPLs  whose 
rightmost index is 0. It turns out, however, that it is not really necessary 
to work out explicitly also this last case (which is more complicated, 
as also the $-\infty<x\leq 0$ cut is present). One can in fact exploit the 
reduction formulae of the previous section for choosing a set 
of irreducible HPLs  with the rightmost index always different from 0, 
(with the exception of the $w=1$ case, where the irreducible set coincides 
with the set of all 3 HPLs), and then express 
all the HPLs  with $w>1$ and rightmost index 0 in terms of those 
irreducible HPLs. 

It is clear from the above discussion that any {\hpl} of arbitrary weight, 
whose rightmost index is 1 or $-1$ and whose other indices are otherwise 
arbitrary, say $\H(\vec a;x)$, can always be separated into the sum of 
two functions, 
\begin{equation} 
   \H(\vec a;x) = \Hp(\vec a;x) + \Hm(\vec a;x) \ , 
\label{eq:splitw} 
\end{equation} 
where $ \Hp(\vec a;x) $ has only the right cut $ 1\leq x<+\infty $ 
and $ \Hm(\vec a;x) $ only the left cut $ -\infty<x\leq -1 $. 
(The formula applies of course even when one of the two cuts is missing, 
as in that case the corresponding function is equal to zero; one can 
write for instance $ \H(0,1,1;x) = \Hp(0,1,1;x) $, as 
$ \Hm(0,1,1;x) = 0 $.) 
All the involved functions (the HPLs  as well as the functions 
corresponding to the separated cuts with branch points at 1 or $-1$) 
are regular at $x=0$ and can be expanded in series of powers of $x$, 
the constant term always being zero. 
\par 
Algorithmically, if $ {\vec a} = (a,\vec b) $ ({\it i.e.} $a$ is the 
leftmost index of $\vec a$, and $\vec b$ stands for the remaining 
$w-1$ indices) and $\H(\vec b;x)$ admits the separation of the cuts 
\begin{equation} \H(\vec b;x) = \Hp(\vec b;x) + \Hm(\vec b;x) \ , 
\end{equation} 
with the two functions $ \Hp(\vec b;x) $ and $ \Hm(\vec b;x) $ 
corresponding to the right $ 1\leq x<+\infty $ and left 
$ -\infty<x\leq -1 $ cuts, 
the explicit separation formulae for $\H(a,\vec b;x)$ of 
weight $w+1$ 
$$ \H(a,\vec b;x) = \Hp(a,\vec b;x) + \Hm(a,\vec b;x) $$ 
in the three cases $a=1,0,-1$ are given by 
\begin{eqnarray} 
 \Hp(1,\vec b;x)&=& \int_0^x \frac{dt}{1-t} \left[ \Hp(\vec b;x) 
                   + \Hm(\vec b;1) \right] \ , \nonumber\\ 
 \Hm(1,\vec b;x)&=& \int_0^x \frac{dt}{1-t} \left[ \Hm(\vec b;x) 
                   - \Hm(\vec b;1) \right] \ , \nonumber\\ 
 \Hp(0,\vec b;x)&=& \int_0^x \frac{dt}{t}\; \Hp(\vec b;x) \ , \nonumber\\ 
 \Hm(0,\vec b;x)&=& \int_0^x \frac{dt}{t}\; \Hm(\vec b;x) \ , \nonumber\\ 
 \Hp(-1,\vec b;x)&=& \int_0^x \frac{dt}{1+t} \left[ \Hp(\vec b;x) 
                   - \Hp(\vec b;-1) \right] \ , \nonumber\\ 
 \Hm(-1,\vec b;x)&=& \int_0^x \frac{dt}{1+t} \left[ \Hm(\vec b;x) 
                   + \Hp(\vec b;-1) \right] \ . 
\label{eq:split} 
\end{eqnarray} 
The separation of the two cuts is illustrated with two explicit examples: 
\begin{eqnarray} 
 \H(-1,1,1;y) &=& \Hp(-1,1,1;y) + \Hm(-1,1,1;y) \ ,          \nonumber \\ 
\mbox{with:}\quad \Hp(-1,1,1;y) &=& \H(-1,1,1;y) - \frac{1}{2}\,\ln^22\ 
                                                \H(-1;y) \ , \nonumber \\
\Hm(-1,1,1;y) &=& \frac{1}{2}\,\ln^22 \ \H(-1;y) \ , 
\label{eq:split-111} 
\end{eqnarray} 
and 
\begin{eqnarray} 
  \H(0,1,-1,1;x) &=&  \Hp(0,1,-1,1;x) + \Hm(0,1,-1,1;x) \ , \nonumber \\ 
\mbox{with:}\quad  \Hp(0,1,-1,1;x) &=& \H(0,1,-1,1;x) - \ln2\ \H(0,1,-1;x) 
                                    + \ln^22\ \H(0,1;x) \ , \nonumber \\ 
 \Hm(0,1,-1,1;x) &=& \ln2\  \H(0,1,-1;x) - \ln^22\ \H(0,1;x) \ . 
\label{eq:split01-11} 
\end{eqnarray} 

The actually chosen irreducible HPLs  are listed in Table~\ref{tab:irred}.
Note that, in addition to the requirement of the absence of the index 0 
in the rightmost place, one can also impose the absence of the index 1 
in the leftmost place. That guarantees, owing to the definition 
Eq.\ (\ref{eq:defn0}), that all the HPLs  of the above irreducibe set 
have finite values at $x=1$. The explicit  values at $x=1$ are needed 
in the separation of cuts, as well as in some of the transformation 
formulae discussed in the following section (as a matter of fact, the 
$x=1$ values can also be obtained as consistency conditions of the many 
transformation formulae of various kinds, which can be easily 
established for the HPLs). Up to weight 4, they can easily be obtained 
by using the table of definite integrals given in \cite{oneval}, and we 
list them in Table~\ref{tab:irred} as well. It can be seen that, besides 
$\ln 2 = 0.693147180559945\ldots$ and $\pi^2 = 9.869604401089359\ldots$,
only two more transcendental constants appear: 
$\zeta_3 = 1.202056903159594\ldots$ and ${\rm Li}_4(1/2) = 
0.517479061673899\ldots$.
The values at $x=-1$ can be obtained directly from the transformation 
discussed in Section~\ref{subsec:min} below.

\section{Identities for HPLs  of related arguments} 
\label{sec:ids} 
\setcounter{equation}{0} 
We discuss here the identities valid for HPLs  whose arguments are related 
by the transformations $ x = - y $, $ x = 1/t $ and $ x = (1-r)/(1+r) $. 

\subsection{ The $ x=-y $ transformation } 
\label{subsec:min}
For real $x$ (in the whole range $ -\infty<x<+\infty$), if $ y = -x $, 
at weight $w=1$ one has from Eq.\ (\ref{eq:cmp1}) 
\begin{eqnarray} 
  \H(1;x+\iep)  &=& - \H^*(-1;y+\iep) \ , \nonumber\\ 
  \H(0;x+\iep)  &=& \bigl(\; \H(0,y+\iep) - i\pi \;\bigr)^* \ , \nonumber\\ 
  \H(-1;x+\iep) &=& - \H^*(1;y+\iep)  \ , 
\label{eq:xtoy} 
\end{eqnarray} 
where the asterisk stands for the complex conjugate. 
\par 
At $ w>1 $, if the rightmost index is different from 0 (as is the case 
for the irreducible HPLs), writing explicitly the indices up to $w=4$ 
one finds immediately the following formulae, which apply for $a_1=1,-1$ and
any value of the other indices: 
\begin{eqnarray} 
  \H(a_2,a_1;x+\iep) &=& (-1)^{a_1+a_2}  \H^*(-a_2,-a_1;y+\iep) 
                                         \ , \nonumber\\ 
  \H(a_3,a_2,a_1;x+\iep) &=& (-1)^{a_1+a_2+a_3} 
                   \H^*(-a_3,-a_2,-a_1;y+\iep) \ , \nonumber\\ 
  \H(a_4,a_3,a_2,a_1;x+\iep) &=& (-1)^{a_1+a_2+a_3+a_4} 
              \H^*(-a_4,-a_3,-a_2,-a_1;y+\iep) \ . 
\label{eq:xtoy2} 
 \end{eqnarray} 
\subsection{ The $ x=1/t $ transformation } 
\label{subsec:x}
For real $x$ in the range $ 1\leq x<+\infty $, put 
\begin{equation} 
  x = \frac{1}{t} \ , {\hskip1cm} t = \frac{1}{x} \ , 
      {\hskip15mm} 1\leq x<+\infty \ , {\hskip4mm} 1\geq t>0 \ . 
\label{eq:xtot} 
\end{equation} 
At weight 1, one has 
\begin{eqnarray} 
  \H(1;x+\iep) &=& \H(1;t) + \H(0,t) + i\pi \ , \nonumber\\ 
  \H(0;x)      &=& - \H(0,t)                \ , \nonumber\\ 
  \H(-1;x)     &=& \H(-1;t) - \H(0,t) \ . 
\label{eq:xtotw1} 
\end{eqnarray} 
For higher weight, one can proceed recursively ({\ie} by 
induction on the weight $w$), observing further that 
for $w > 1$ it is sufficient to establish the identities for the 
irreducible HPLs  only, as the other cases can be obtained through the 
reduction formulae. Let us consider a generic irreducible \hpl, 
say $\H(a,\vec b;x=1/t)$, and define accordingly 
$$ X(a,\vec b;t) = \H(a,\vec b;1/t) \ . $$ 
Quite in general, 
$$ X(a,\vec b;t) = X(a,\vec b;1) + \int_1^t dt' 
                   \frac{d}{dt'} X(a,\vec b;t') \ ; $$ 
by using, in the r.h.s., the very definition of $ X(a,\vec b;t) $ and 
Eq.\ (\ref{eq:defn0}) the previous equation reads 
$$ \H\left(a,\vec b;\frac{1}{t}\right) = \H(a,\vec b;1) 
           - \int_1^t dt' \frac{1}{t'^2} \; 
            \f\left(a,\frac{1}{t'}\right) \; 
            \H\left(\vec b;\frac{1}{t'}\right) \ . $$ 
It was already remarked that the irreducible HPLs  can be chosen with 
the leftmost index different from 1; therefore in the above equation 
the index $a$ can take only the values 0 and $-1$; correspondingly 
one finds 
\begin{eqnarray} 
   \H\left(0,\vec b;\frac{1}{t}\right) &=& \H(0,\vec b;1) 
                   - \int_1^t dt'\;\frac{1}{t'} \; 
                   \H\left(\vec b;\frac{1}{t'}\right) \ , \nonumber\\ 
   \H\left(-1,\vec b;\frac{1}{t}\right) &=& \H(-1,\vec b;1) 
          - \int_1^t dt'\; \left( \frac{1}{t'} - \frac{1}{1+t'} \right) 
                \; \H\left(\vec b;\frac{1}{t'}\right) \ . 
\label{eq:xtotww} 
\end{eqnarray} 
By proceeding recursively from lower to higher weights, $ \H(\vec b;1/t') $ 
can be considered as already given in terms of HPLs  of argument 
$t'$, so that due to Eq.\ (\ref{eq:defn0}) $ X(a,\vec b;t) $, hence 
$ \H(a,\vec b;x=1/t) $ is also expressed in terms of HPLs  of argument 
$t$. 

As an example of this transformation, one finds for instance, 
\begin{eqnarray}
\H(-1,1;x) & = & -\H(0,1;t) - \H(0,-1;t) + \H(-1,1;t) - \frac{1}{2}
                \H(0;t)\H(0;t) + \H(-1;t)\H(0;t) + \frac{\pi^2}{4} \nonumber\\
&& + i\pi \biggl( \H(-1;t) - \H(0;t) - \ln 2\biggr)\ , \nonumber \\
\nonumber \\
\H(0,1,1;x) & = & \H(0,0,1;t) - \H(0,1,1;t) 
                + \H(0;t)\biggl( \frac{\pi^2}{2} - \H(0,1;t) - \frac{1}{6}
         \H(0;t)\H(0;t) \biggr) + \zeta_3 \nonumber \\
           && + i\pi \biggl(- \H(0,1;t) - \frac{1}{2}
         \H(0;t)\H(0;t) + \frac{\pi^2}{6}  \biggr)\ .
\end{eqnarray}

\subsection{ The $ x=(1-r)/(1+r) $ transformation } 
\label{subsec:r}
The transformation 
\begin{equation} 
 x = \frac{1-r}{1+r}\ , {\hskip15mm} r = \frac{1-x}{1+x}\ \ , 
 {\hskip10mm} 0 \leq x < +\infty \ , {\hskip4mm} 1 \geq r > -1 \ , 
\label{eq:xtor} 
\end{equation} 
can be treated similarly. For $ x $ in the range $ 0\leq x\leq 1 $, $ r $ 
is positive, $ 1\geq r\geq 0 $, and all the involved functions are real, 
while for $x$ in the range $ 1\leq x<+\infty$ the variable $r$ is negative, 
$ 0\geq r>-1 $; at weight $w=1$, therefore, one has 
\begin{eqnarray} 
 \H(1;x)  &=& - \H(0;|r|) + \H(-1;r) - \ln2 +i\pi\theta(-r) \ , \nonumber\\ 
 \H(0;x)  &=& - \H(1;r) - \H(-1;r)        \ , \nonumber\\ 
 \H(-1;x) &=& - \H(-1;r) + \ln2 \ .
\label{eq:xtorw1} 
\end{eqnarray} 
To work out the corresponding formulae for any of the irreducible {\hpl} 
of higher weight, say $\H(a,\vec b;x=(1-r)/(1+r))$, define 
$$ Y(a,\vec b;r) = \H\left(a,\vec b;\frac{1-r}{1+r}\right) \ , $$ 
and evaluate it through the obvious relation 
$$ Y(a,\vec b;r) = Y(a,\vec b;0) + \int_0^r dr' \frac{d}{dr'} 
                   Y(a,\vec b;r') \ , $$ 
which on account of the definitions becomes 
$$ \H\left(a,\vec b;\frac{1-r}{1+r}\right) = \H(a,\vec b;1) 
                 - \int_0^r dr' \frac{2}{(1+r')^2} 
                   \f\left(a,\frac{1-r'}{1+r'}\right) 
                   \H\left(\vec b;\frac{1-r'}{1+r'}\right) \ . $$ 
It was already recalled that the leftmost index of an irreducible 
{\hpl} does not take the value 1; for $a=0,-1$ an elementary explicit 
calculation gives 
\begin{eqnarray} 
  \H\left(0,\vec b;\frac{1-r}{1+r}\right) &=& \H(0,\vec b;1) 
          - \int_0^r dr' \left( \frac{1}{1-r'} 
       + \frac{1}{1+r'} \right) \H\left(\vec b;\frac{1-r'}{1+r'}\right) 
                                                        \ , \nonumber\\ 
  \H\left(-1,\vec b;\frac{1-r}{1+r}\right) &=& \H(-1,\vec b;1) 
       - \int_0^r dr'\; \frac{1}{1+r'} 
       \H\left(\vec b;\frac{1-r'}{1+r'}\right) \ . 
\label{eq:xtorww} 
\end{eqnarray} 
Proceeding again recursively from lower to higher weights, 
$ \H(\vec b;(1-r')/(1+r')) $ 
can be considered as already given in terms of HPLs  of argument 
$r'$; therefore due to Eq.\ (\ref{eq:defn0}) $ Y(a,\vec b;r) $, hence 
$ \H(a,\vec b;x=(1-r)/(1+r)) $, is also expressed in terms of HPLs  
of argument $r$. 
\par 
As an example of the above identities, written in a form that applies 
in the whole range $0<x<+\infty$ one finds for instance 
\begin{eqnarray} 
  \H(0,1;x) &= &- \H(0,1;r) - \H(0,-1;r) + \H(-1,1;r) \nonumber\\ 
            &&+ \biggl( \H(1;r) + \H(-1;r) \biggr) 
                \biggl( \H(0;|r|) + \ln2 - i\pi\theta(-r) \biggr) \nonumber\\ 
            &&- \left( \H(1;r) + \frac{1}{2}\H(-1;r) \right) \H(-1;r) 
                       + \frac{\pi^2}{6} \ ,\nonumber \\
  \H(-1,1;x)&= &-\H(0,-1;r) + \H(-1;r) \biggl( \H(0;|r|)+\ln2 
                                 - i\pi\theta(-r) \biggr) 
                         \nonumber \\
            && - \frac{1}{2} \H(-1;r)\H(-1;r) + \frac{\pi^2}{12} - 
                            \frac{\ln^2 2}{2} \ .
\end{eqnarray} 

As a last remark concerning the transformation (\ref{eq:xtor}), let us 
observe that its fixed points satisfy the equation 
$$ x = \frac{1-x}{1+x} \ , $$ 
whose solutions are 
\begin{eqnarray} 
  x_+ &=& +\; (\sqrt{2} - 1 ) \ , \nonumber\\ 
  x_- &=& -\; (\sqrt{2} + 1 ) \ . 
\label{eq:r2pm1} 
\end{eqnarray} 

\section{The numerical evaluation} 
\label{sec:numer} 
\setcounter{equation}{0} 
The analytical properties discussed in the previous 
sections can be exploited to get a fast and precise numerical evaluation 
(absolute error less than $3\times 10^{-15}$) of the HPLs  of real but 
otherwise arbitrary argument $x$. We divide the whole range 
$-\infty<x<+\infty$ into the 5 regions 
\begin{itemize} 
\item $ -\infty<x < -(\sqrt{2}+1) $, 
\item $ -(\sqrt{2}+1)\leq x < -(\sqrt{2}-1) $, 
\item $ -(\sqrt{2}-1)\leq x \leq +(\sqrt{2}-1) $, 
\item $(\sqrt{2}-1)<x   \leq (\sqrt{2}+1)$, 
\item $ (\sqrt{2}+1)<x<+\infty $. 
\end{itemize} 
In evaluating all HPLs, 
it is always assumed that the (infinitesimal) imaginary part of
the argument  is positive, {\it i.e.}\ the HPLs  are evaluated for 
argument $x+i\epsilon$. 

All transformation and expansion formulae used in the numerical 
evaluation of the HPLs were obtained using the computer algebra
program FORM~\cite{form}.

\par 
\subsection{The central region $ -(\sqrt{2}-1)\leq x\leq+(\sqrt{2}-1) $ } 
In the central region $ -(\sqrt{2}-1)\leq x\leq+(\sqrt{2}-1) $, we 
evaluate $ \H(0;x) $, according to its definition Eq.\ (\ref{eq:defineh1}), 
as $ \ln(|x|) + i\pi\theta(-x) $, and all the other irreducible HPLs  
up to the required weight through a power series expansion in $x$ 
around $x=0$.

As $ |x| \le (\sqrt{2}-1) = 0.4142\ldots$, 
to obtain the aimed $3\times 10^{-15}$ 
absolute precision, one should keep in principle
 34 terms in the expansion in powers of $x$. If the indices of some 
HPLs  take only the values 0 and 1, it has already been observed that 
the corresponding function has only the right cut $ 1\leq x<+\infty $; in 
that case, the convergence of 
the power series expansions improves considerably by 
the Bernoulli~\cite{btrans} change of variable 
\begin{equation} 
  x = 1 - \e^{-u} \ , {\hskip15mm} u = - \; \ln(1-x) \ .
\label{eq:Bernu} 
\end{equation} 
This transformation 
is invertible in the strip $ |\Im u| < 2\pi $ and moves the $ x=1 $ 
branch point to $ u = +\infty $. By expanding in powers of $u$ at 
$u=0$ one obtains a series in $u$ whose radius of convergence is $2\pi$ 
(as opposed to 1 for the series in $x$). As $x$ moves from 
$ -(\sqrt{2}-1) $ through 0 to $ (\sqrt{2}-1) $, $ u/(2\pi) $ 
moves from $ -0.055\ldots $ through 0 to $ 0.085\ldots$, so that in the worst 
case $ |u/(2\pi)| \sim 0.085\ldots $; 
to attain a $10^{-16}$ precision it is sufficient to keep up 
to 15 terms in $u$. A few terms drop when using standard Chebyshev 
economization, so that in practice the required accuracy is obtained 
with at most 12 Chebyshev polynomials in $u$. The same approach works 
when the indices of the HPLs  take only the values 0 and $-1$; in that 
case the corresponding function has only the left cut 
$ -\infty<x\leq -1 $ and the appropriate change of variable is 
\begin{equation} 
  x = \e^{v}-1 \ , {\hskip15mm} v =  \ln(1+x) \ ; 
\label{eq:Bernv} 
\end{equation} 
it is found again that after Chebyshev economization at most 12 Chebyshev 
polynomials in $v$ are needed.

When 1 and $-$1 both appear as indices of the HPLs  both cuts are 
present. An expansion in powers of $x$ at $x=0$ can be equally 
carried out, but the above changes of variable do not help for speeding 
up convergence. Indeed, by using for instance the transformation 
Eq.\ (\ref{eq:Bernu}), the left branch point at $x=-1$ is mapped into 
$ v = - \ln2 $, so that the radius of convergence of the 
$u$ expansion at $u=0$ would be in fact reduced from 1 to 
$\ln2$ ( $\ln2 = 0.6931\ldots $). 
This problem is, however, easily overcome by using Eq.\ (\ref{eq:splitw}) 
and evaluating independently, for each irreducible $\H(\vec a;x)$, the 
two related functions $\Hp(\vec a;x)$ and $\Hm(\vec a;x)$ by  
using the variable $u$ of Eq.\ (\ref{eq:Bernu}) for $\Hp(\vec a;x)$ 
and the variable $v$ of Eq.\ (\ref{eq:Bernv}) for $\Hm(\vec a;x)$. 
It should be noted that it is not actually necessary to evaluate both 
$\Hp(\vec a;x)$ and $\Hm(\vec a;x)$ for each irreducible $\H(\vec a;x)$. 
Indeed, if the rightmost index is 1, 
$\Hp(\vec a;x)$ differs from $\H(\vec a;x)$ by a combination of 
constants times HPLs  with the left cut and smaller weight, 
see for instance Eqs.\ (\ref{eq:split-111}) and (\ref{eq:split01-11}), while 
$\Hm(\vec a;x)$ is just that difference. 
In a systematic approach to the numerical evaluation of the HPLs  
in order of increasing weight, such lower weight functions can be 
considered as already evaluated. Therefore only $\Hp(\vec a;x)$ has 
to be evaluated from scratch as a power series in the suitable variable. 
The obvious analogue applies when the rightmost index is $-1$. 

Some explicit examples of actual expansions in $u$ and $v$ follow:
\begin{eqnarray}
\H(0,1;x) & = & u - \frac{u^2}{4} + \frac{u^3}{36} - \frac{u^5}{3600}
+ \frac{u^7}{211680} - \frac{u^9}{10886400} + \frac{u^{11}}{526901760}
- \frac{691u^{13}}{16999766784000} + \ldots \nonumber \\
\H(-1,1;x) & = & - \ln 2 \; \H(-1;x)  \nonumber \\ && 
     + u\ \ln 2 + \left( \frac{1}{2} - \ln 2\right) u^2
                          + \left(-\frac{2}{3} + \ln 2\right) u^3
     + \left( \frac{3}{4} -\frac{13}{12}\ \ln 2\right) u^4
     + \left(-\frac{13}{15} +\frac{5}{4}\ \ln 2\right) u^5 \nonumber \\ &&
     + \left( \frac{25}{24} -\frac{541}{360}\ \ln 2\right) u^6
     + \left(-\frac{541}{420} +\frac{223}{120}\ \ln 2\right) u^7
     + \left( \frac{1561}{960} -\frac{47293}{20160}\ \ln 2\right) u^8
\nonumber \\ &&
     + \left(-\frac{47293}{22680} +\frac{36389}{12096}\ \ln 2\right) u^9
     + \left( \frac{36389}{13440} -\frac{7087261}{1814400}\ \ln 2\right) u^{10}
\nonumber \\ &&
     + \left(-\frac{7087261}{1995840} +\frac{3098411}{604800}\ \ln 2\right) u^{11}
     + \left( \frac{34082521}{7257600} -\frac{1622632573}{239500800}\ \ln 2\right) u^{12}\nonumber \\ &&
     + \left(-\frac{1622632573}{259459200} +\frac{20579903}{2280960}\ \ln 2\right) u^{13}
+ \ldots
\nonumber \\
\H(0,0,-1;x) & = & v + \frac{3v^2}{8} + \frac{17v^3}{216} 
           + \frac{5v^4}{576} + \frac{7v^5}{54000} - \frac{7v^6}{86400}
- \frac{19v^7}{ 5556600} + \frac{v^8}{752640} + \frac{11v^9}{127008000}
\nonumber \\ &&
- \frac{11v^{10}}{435456000} - \frac{3263v^{11}}{1521428832000} + 
\frac{13v^{12}}{25291284480} + \frac{13399637v^{13}}{255251498261760000}
+ \ldots 
\end{eqnarray} 
The fast convergence of the expansion is shown by the fast decrease 
of the higher order coefficients. In the first and third of the above 
equations the fast decrease is manifest; in the second equation it comes 
from the strong cancellations between the two terms in brackets; 
for instance, the actual value of the coefficient of the $u^4$ term is 
$ 9.09\ldots\times 10^{-4} $, the first rational fraction being $3/4$ 
(the coefficient could indeed be used for obtaining the 
rational approximation $9/13 = 0.692307... $ to $ \ln2 = 0.6923147... $), 
and the effect is of course enhanced in the subsequent terms. 
Some care has therefore to be taken in implementing these coefficients in a 
numerical program. To avoid large cancellations inside the coefficients 
during the execution of the program, we converted all coefficients from 
combinations of rational fractions and transcendental constants into 
real numbers of the desired accuracy, so that in the {\tt FORTRAN} 
code each power of $u$ and $v$ is multiplied by a single constant in 
double precision. 

It should be noted that $u$ and $v$ are both in the range 
$[-\ln(\sqrt{2}+2):\ln(\sqrt{2}+2)]$.
The Chebyshev economization is carried out on those terms by rescaling 
$u$ and $v$ with a factor $20/11$, to an interval slightly smaller than 
$[-1:1]$. Using this only approximate rescaling allows an 
extension of the 
numerical formulae used in this region also slightly beyond its 
boundaries, which will be used as a check on the numerical program below.

\subsection{ The region  $ (\sqrt{2}-1)<x\leq(\sqrt{2}+1) $ } 
\label{subsec:expandr}
In the region $ (\sqrt{2}-1)<x\leq(\sqrt{2}+1) $ we use the change 
of variable of Eq.\ (\ref{eq:xtor}) in Section~\ref{subsec:r} 
for expressing the irreducible HPLs  of argument $x$ in terms of the 
irreducible HPLs  of argument $r$, with 
$$ r = \frac{1-x}{1+x} \ . $$ 
Note that $ x=(\sqrt{2}-1) $ corresponds to $ r = (\sqrt{2}-1) $ (one 
of the fixed points of the transformation), while $ x=(\sqrt{2}+1) $ 
corresponds to $ r = - (\sqrt{2}-1) $, so that the region 
$ (\sqrt{2}-1)<x\leq(\sqrt{2}+1) $ is mapped into the $r$-range 
$ - (\sqrt{2}-1) \leq r < (\sqrt{2}-1) $, which is exactly the 
central region discussed in the previous subsection. We 
evaluate the HPLs  of $r$ with the series expansion discussed above,
then obtain the required HPLs  of $x$ by means of the transformation 
formulae discussed in Section~\ref{subsec:r}.
\subsection{ The region  $ (\sqrt{2}+1)<x<+\infty $ }
\label{subsec:expandx} 
We use the change of variable $x=1/t$ of Eq.\ (\ref{eq:xtot}), discussed 
in Section~\ref{subsec:x}, for expressing the irreducible HPLs  of 
argument $x$ in terms of the irreducible HPLs  of argument $t$, with
$ t = 1/x $. Note that $ x = (\sqrt{2}+1) $ corresponds to 
$ t = (\sqrt{2}-1) $, while $ x \to +\infty $ translates
 into $ t\to 0 $, so that 
the region $ (\sqrt{2}+1)<x<+\infty $ is mapped into the $t$-range 
$ 0 < t < (\sqrt{2}-1) $, which is the positive half of the 
central region discussed at the beginning of this section. We 
evaluate the HPLs  of $t$ from their series expansion, 
then obtain the required HPLs  of $x$ by means of the transformation 
formulae of Section~\ref{subsec:x}.
\subsection{ The region  $ -(\sqrt{2}+1)\leq x<-(\sqrt{2}-1) $ } 
We use the change of variable $x=-y$. We evaluate the HPLs  of 
argument $ y = -x $ as in Subsection~\ref{subsec:expandr} and then use 
the formulae of Subsection~\ref{subsec:min} for expressing the required 
HPLs  of argument $x$ in terms of the HPLs   of argument $y$. 
\subsection{ The region  $ -\infty<x<-(\sqrt{2}+1) $ } 
Essentially the same as before. We evaluate the HPLs  of argument 
$ y = -x $ as in Subsection~\ref{subsec:expandx} and then use the 
formulae of Section~\ref{subsec:min}.

\section{Checks}
\label{sec:checks}
We have carried out several checks on our implementation of the 
algorithm described in the previous section into a {\tt FORTRAN} 
subroutine. 

An immediate check of the numerical implementation of the HPLs  
is provided
 by the derivative formula Eq.\ (\ref{eq:derive}).
Evaluating the left-hand side of Eq.\ (\ref{eq:derive}) numerically
with a standard symmetric 4-point differentiation formula, and comparing 
it with the right-hand side evaluated directly, we found agreement within 
an accuracy of $10^{-12}$ or better. This accuracy is mainly limited by 
rounding errors induced by the small interval size used in the 
differentiation formula. We used $10^{-4}$ as interval size, which 
implies a theoretical accuracy of about $10^{-16}$. This accuracy is 
however reduced to the observed $10^{-12}$ by rounding errors arising 
from taking the difference between the function values evaluated 
at interval points spaced by $10^{-4}$ only. 

We also checked the continuity of all HPLs  across the boundaries of 
the different regions introduced in Section~\ref{sec:numer}, which match onto 
each other at the points $\pm x_+$ and $\pm x_-$ of Eq.\ (\ref{eq:r2pm1}).
Evaluating the HPLs  according to the algorithms appropriate to the 
regions left and right of the boundaries, we found both limiting 
values to agree within $3\times 10^{-15}$ or better for 
both real and imaginary parts. Moreover, we evaluated 
the HPLs  in a few points scattered in a small neighbourhood (of size 
$\pm 10^{-10}$) across the boundaries of the regions discussed in the 
previous section, by using for each point the two different algorithms 
used separately in the {\tt FORTRAN} code at each side of the boundaries 
and then comparing the results. Again, we found agreement within the 
desired accuracy of $3\times 10^{-15}$. 

\section{The subroutine {\tt hplog}}
\label{sec:code}

\subsection{Syntax}
The routine {\tt hplog} has the following syntax:
\begin{verbatim}
      subroutine hplog(x,nw,Hc1,Hc2,Hc3,Hc4,
     $                      Hr1,Hr2,Hr3,Hr4,Hi1,Hi2,Hi3,Hi4,n1,n2)
\end{verbatim}

\subsection{Usage}
In calling {\tt hplog}, the user has to supply 
\begin{itemize}
\item[{{\tt x}:}] The argument for which the HPLs  are to be evaluated.
{\tt x} is of type {\tt real*8}. It can take any real value 
$-\infty < x < \infty$. 
\item[{{\tt nw}:}] The maximum weight of the  HPLs  to be evaluated.
{\tt nw} is of type {\tt integer}. It is limited to $1\leq {\tt nw} \leq 4$. 
\item[{{\tt n1,n2}:}] Define the indices for which the HPLs  are 
evaluated. {\tt(n1,n2)} are of type {\tt integer}. Allowed combinations are 
$(-1,1)$ (evaluate all HPLs), $(0,1)$ (evaluate only HPLs  with 
0 and 1 appearing in the vector of arguments) and $(-1,0)$
(evaluate only HPLs  with 
0 and $-1$ appearing in the vector of arguments).
\end{itemize}

The output of {\tt hplog} is provided in the arrays 
{\tt Hc1,Hc2,Hc3,Hc4,Hr1,Hr2,Hr3,Hr4,Hi1,Hi2,Hi3,Hi4}. These have to be 
declared and dimensioned by the user as follows:
\begin{verbatim}
      complex*16 Hc1,Hc2,Hc3,Hc4 
      real*8     Hr1,Hr2,Hr3,Hr4 
      real*8     Hi1,Hi2,Hi3,Hi4 
      dimension Hc1(n1:n2),Hc2(n1:n2,n1:n2),Hc3(n1:n2,n1:n2,n1:n2), 
     $          Hc4(n1:n2,n1:n2,n1:n2,n1:n2) 
      dimension Hr1(n1:n2),Hr2(n1:n2,n1:n2),Hr3(n1:n2,n1:n2,n1:n2), 
     $          Hr4(n1:n2,n1:n2,n1:n2,n1:n2) 
      dimension Hi1(n1:n2),Hi2(n1:n2,n1:n2),Hi3(n1:n2,n1:n2,n1:n2), 
     $          Hi4(n1:n2,n1:n2,n1:n2,n1:n2) 
\end{verbatim} 
It should be noted that this declaration is always needed, even if
{\tt hplog} is called with {\tt nw}$<4$. After calling {\tt hplog} for 
a given argument {\tt x}, the complex arrays {\tt Hc1, Hc2, ...} contain the 
complex values of the corresponding HPLs  of weight $1,2,\ldots$,
the real arrays {\tt Hr1, Hr2, ...} contain the real values 
and {\tt Hi1, Hi2, ...} the immaginary parts divided by $\pi$, 
so that, for instance, the following equality holds 
\begin{verbatim}
      Hc3(k1,k2,k3) = cmplx( Hr3(k1,k2,k3), pi*Hi3(k1,k2,k3) )
\end{verbatim}
The complex parts of the HPLs  are always evaluated 
assuming an infinitesimal positive imaginary part of the argument.

The subroutine does not need initialization.

\subsection{Example}
The following example program illustrates how to evaluate 
HPLs  up to weight 2 for a given value of $x$, and to write out their 
real parts:
\begin{verbatim}
      program testhpl
      integer n1,n2,nw,i1,i2
      parameter (n1=-1) 
      parameter (n2=1) 
      parameter (nw=2)
      complex*16 Hc1,Hc2,Hc3,Hc4 
      real*8     Hr1,Hr2,Hr3,Hr4 
      real*8     Hi1,Hi2,Hi3,Hi4 
      real*8     x
      dimension Hc1(n1:n2),Hc2(n1:n2,n1:n2),Hc3(n1:n2,n1:n2,n1:n2), 
     $          Hc4(n1:n2,n1:n2,n1:n2,n1:n2) 
      dimension Hr1(n1:n2),Hr2(n1:n2,n1:n2),Hr3(n1:n2,n1:n2,n1:n2), 
     $          Hr4(n1:n2,n1:n2,n1:n2,n1:n2) 
      dimension Hi1(n1:n2),Hi2(n1:n2,n1:n2),Hi3(n1:n2,n1:n2,n1:n2), 
     $          Hi4(n1:n2,n1:n2,n1:n2,n1:n2) 
      write(6,*) 'Input x:' 
      read(5,*) x 
      call hplog(x,nw,Hc1,Hc2,Hc3,Hc4, 
     $                Hr1,Hr2,Hr3,Hr4,Hi1,Hi2,Hi3,Hi4,n1,n2) 
      do i1 = n1,n2
         write(6,101) i1,Hr1(i1)
         do i2 = n1,n2 
            write(6,102) i1,i2,Hr2(i1,i2)
         enddo
      enddo
      stop
  101 format('       H(',i2,',x) = ',f18.15)
  102 format('       H(',i2,',',i2,',x) = ',f18.15)
      end
\end{verbatim}

\section{Numerical examples}
\label{sec:plots}
\begin{figure}[thb]
\begin{center}
~\epsfig{file=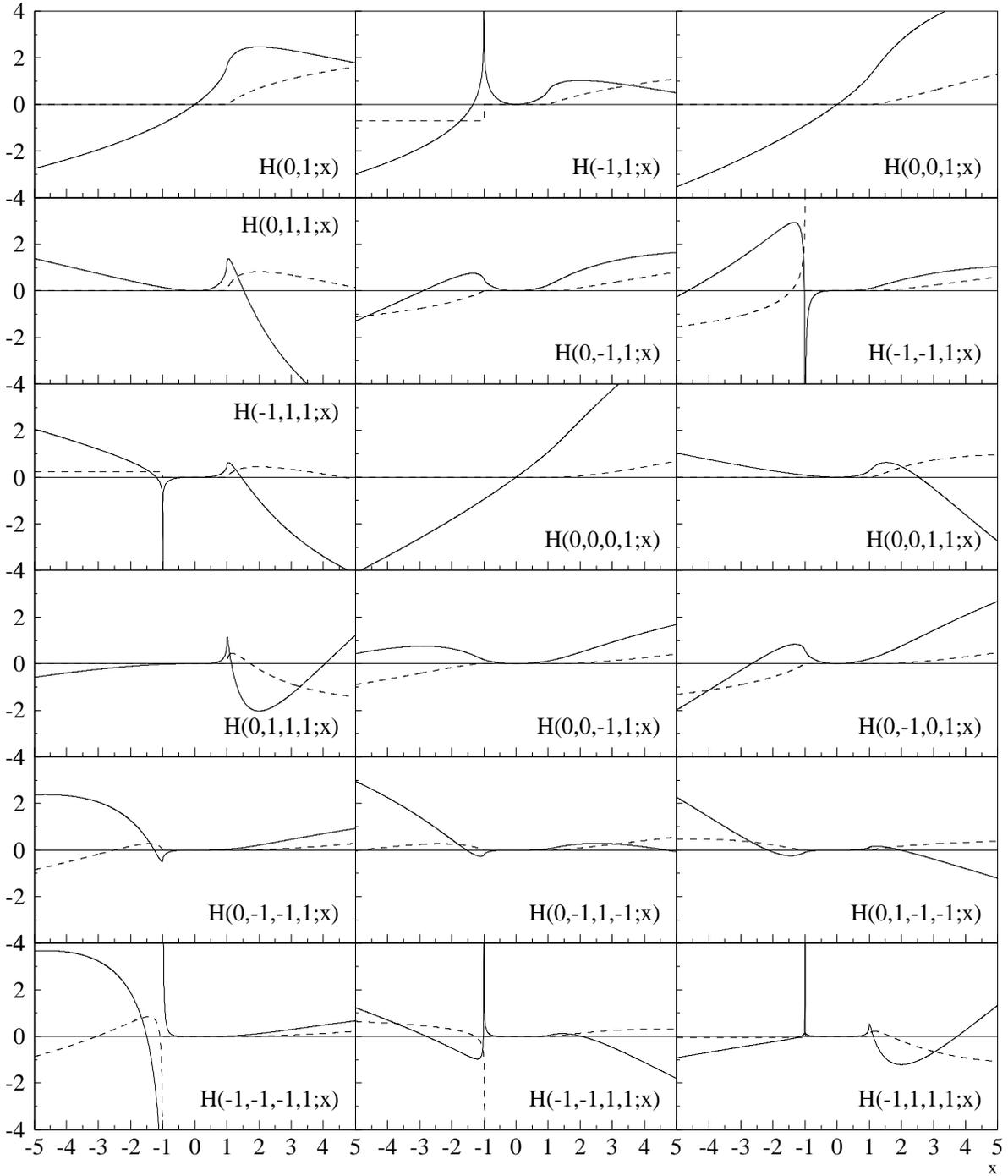,angle=-90,width=16cm}
\label{fig:hpl}
\caption{Plots of irreducible HPLs  in the interval $[-5;5]$. 
Irreducible HPLs  
that follow from the above by $x\to -x$ are left out. Solid line: real
part, dashed line: imaginary part.}
\end{center}
\end{figure}

In Fig.~\ref{fig:hpl}, we plot real and imaginary parts for 
a selection of irreducibe HPLs
over the interval $[-5;5]$. The remaining irreducible HPLs  can be obtained 
by simply interchanging $x\to -x$, according to 
Eqs.\ (\ref{eq:xtoy}),(\ref{eq:xtoy2}).

\section{Summary} 
In this paper, we have described the routine {\tt hplog}, which 
evaluates the harmonic polylogarithms up to weight 4 for 
arbitrary real arguments. The evaluation 
is based on a series expansion in terms of appropriately transformed 
expansion parameters for small values of the argument. The evaluation 
for large arguments is based on transformation formulae, relating 
HPLs  of different arguments. The algorithms used and described here can be
extended to higher weights without further modification, requiring only 
the values of the HPLs  in $x=1$ to be known.  

\section*{Acknowledgement} 
We are grateful to Jos Vermaseren for his assistance in the use of 
the algebraic program FORM~\cite{form}, which was employed intensively  
for generating the code described here.

\newpage
\begin{table}[thb]
\begin{center}
\begin{tabular}{|r|r|r|}\hline
\rule[0mm]{0mm}{5mm}
Weight & HPL & HPL at $x=1$ \\[2mm] \hline 
\rule[0mm]{0mm}{5mm}
       &  $\H(1;x)$ & $\infty$ \\
 $w=1$ & $\H(0;x)$ & 0 \\
& $\H(-1;x)$ & $\ln 2$ \\[2mm] 
\hline 
\rule[0mm]{0mm}{5mm}
 & $\H(0,1;x)$ & $\pi^2/6$ \\ 
 $w=2$ & $\H(0,-1;x)$ & $\pi^2/12$ \\
& $\H(-1,1;x)$ & $\pi^2/12 - \ln^2 2/2$\\[2mm] \hline
\rule[0mm]{0mm}{5mm}
 & $\ \H(0,0,1;x)$ & $\zeta_3$ \\
 & $\ \H(0,1,1;x)$ &  $\zeta_3$ \\
 & $\ \H(0,0,-1;x)$ & $3\zeta_3/4$ \\
 & $\ \H(0,-1,-1;x)$ &  $\zeta_3/8$ \\
\raisebox{1.5ex}[-1.5ex]{$w=3$}
 & $\ \H(0,-1,1;x)$ & $-\pi^2 \ln 2/4 + 13\zeta_3/8$ \\
 & $\ \H(0,1,-1;x)$ & $ \pi^2 \ln 2/4 - \zeta_3$ \\
 & $\ \H(-1,-1,1;x)$ & $ -\ln^3 2/6 + \zeta_3/8$ \\
 & $\ \H(-1,1,1;x)$ & $-\pi^2\ln 2/12 + 7 \zeta_3/8 + \ln^3 2/6$\\[2mm] \hline 
\rule[0mm]{0mm}{5mm}
& $\ \H(0,0,0,1;x)$ & $\pi^4/90$ \\
& $\ \H(0,0,1,1;x)$ & $\pi^4/360$ \\
& $\ \H(0,1,1,1;x)$ & $\pi^4/90$ \\
& $\ \H(0,0,0,-1;x)$ & $7 \pi^4/720$ \\
& $\ \H(0,0,-1,-1;x)$ & $- \pi^2\ln^2 2/12 -\pi^4/48 + 7\zeta_3\ln 2/4 
                                + \ln^4 2/12 + 2 {\rm Li}_4(1/2)$ \\
& $\ \H(0,-1,-1,-1;x)$ & $ \pi^2\ln^2 2/24 +\pi^4/90 - 7\zeta_3\ln 2/8 
                                - \ln^4 2/24 - {\rm Li}_4(1/2)$ \\
& $\ \H(0,0,-1,1;x)$ & $-\pi^2 \ln^2 2 /12 -\pi^4/180 + \ln^4 2/12 
                              + 2 {\rm Li}_4(1/2)$ \\
& $\ \H(0,0,1,-1;x)$ & $-19\pi^4/1440 + 7\zeta_3\ln 2/4$\\
& $\ \H(0,-1,0,1;x)$ & $\pi^4/480$\\
\raisebox{1.5ex}[-1.5ex]{$w=4$}
& $\ \H(0,-1,-1,1;x)$ & $\pi^2\ln^2 2/24 -\pi^4/80 + \ln^4 2/12 
                              + 2 {\rm Li}_4(1/2) $ \\
& $\ \H(0,-1,1,-1;x)$ & $-\pi^2\ln^2 2/4 -7\pi^4/720 +21\zeta_3\ln 2/8$ \\
& $\ \H(0,1,-1,-1;x)$ & $5\pi^2\ln^2 2/24 +7\pi^4/288 -21\zeta_3\ln 2/8
                             - \ln^4 2/12 
                              - 2 {\rm Li}_4(1/2) $ \\
& $\ \H(0,-1,1,1;x)$ & $ -11\pi^4/720 + \ln^4 2/8 +3 {\rm Li}_4(1/2) $ \\
& $\ \H(0,1,-1,1;x)$ & $-\pi^2\ln^2 2/8+7\pi^4/288- \ln^4 2/8 
                               - 3 {\rm Li}_4(1/2) $\\
& $\ \H(0,1,1,-1;x)$ & $\pi^2\ln^2 2/12-\pi^4/80 +7\zeta_3\ln 2/8 
                            +\ln^4 2/24 + {\rm Li}_4(1/2) $ \\
& $\ \H(-1,-1,-1,1;x)$ & $\pi^2\ln^2 2/24+ \pi^4/90 -7\zeta_3\ln 2/8 
                            - \ln^4 2/12 -{\rm Li}_4(1/2) $\\
& $\ \H(-1,-1,1,1;x)$ & $\pi^4/720 -\zeta_3\ln 2/8 +\ln^4 2/24 $ \\
& $\ \H(-1,1,1,1;x)$ &${\rm Li}_4(1/2)$ \\[2mm] \hline
\end{tabular}
\caption{List of irreducible HPLs  chosen 
in the  numerical implementation, and their values for $x=1$.}
\label{tab:irred}
\end{center}
\end{table}

\end{document}